# Low hardware consumption, resolution-configurable gray code oscillator time-to-digital converters implemented in 16nm, 20nm and 28nm FPGAs

Yu Wang, Wujun Xie, Haochang Chen, and David Day-Uei Li

***Abstract*—This paper presents a low-hardware consumption, resolution-configurable, automatically calibrating gray code oscillator time-to-digital converter (GCO-TDC) in Xilinx 16nm UltraScale+, 20nm UltraScale and 28nm Virtex-7 field-programmable gate arrays (FPGAs). The proposed TDC utilizes LUTs as delay elements and has several innovations: 1) a sampling matrix structure to improve the resolution, 2) a virtual bin calibration method (VBCM) to achieve configurable resolutions and automatic calibration, and 3) hardware implementation of the VBCM in standard FPGA devices. We implemented and evaluated a 16-channel TDC system in all three FPGAs. The UltraScale+ version achieved the best resolution (least significant bit, LSB) of 20.97 ps with 0.09 LSB averaged peak-to-peak differential nonlinearity ($DNL_{pk\text{-}pk}$). The UltraScale and Virtex-7 versions achieved the best resolutions of 36.01 ps with 0.10 LSB averaged $DNL_{pk\text{-}pk}$ and 34.84 ps with 0.08 LSB averaged $DNL_{pk\text{-}pk}$, respectively.***

## I. INTRODUCTION

Time-to-digital converters (TDCs) are simply high-precision time-interval meters, converting a time interval (TI) into a digital code. They are widely used in industrial and scientific applications, including light detection and ranging (LiDAR) for automatic vehicles and robotics [1]–[3], 3-D imaging [4]–[6], surveying [7], Raman spectroscopy [8], [9], hardware trojan detection [10], temperature sensing [11], [12], random number generation [13], [14], particle physics [15]–[17], positron emission tomography (PET) [18], fluorescence lifetime imaging microscopy (FLIM) [19] and space sciences [20].

The primary metric of a TDC is the resolution (the minimum TI that can be measured, also called the least significant bit, LSB). Ideally, all bins in a TDC should have the same width. The ideal bin width can be defined as $Q = T/n$, where $T$ is the period of the sampling clock and $n$ is the number of bins, respectively. However, bin widths are not uniform. The variations of bin widths caused by uneven delay elements [21] are usually characterized by differential nonlinearity (DNL) and integrated nonlinearity (INL). They are respectively defined as $DNL[k]=(W[k]-Q)/Q$ and $INL[k] = \sum_{n=0}^{k} DNL[j]$, where $W[k]$ is the $k$-th bin's width, and it can be evaluated by code density tests [22]. With recent advances in CMOS manufacturing technologies, FPGA-TDCs and ASIC-TDCs can achieve picosecond-level resolutions. However, FPGA-TDCs are cheaper and have shorter developing cycles. These characteristics make FPGA-TDCs popular in prototype designs.

Industrial applications utilizing time-of-flight (TOF) information (such as LiDAR) concern not only the resolution but also linearity. In time-resolved LiDAR systems, a resolution of 66.6 ps corresponds to a distance of 1 cm [23]. Therefore, LiDAR systems for automatic vehicles and robotics require TDCs with a 35-500 ps resolution and high linearity [24]. Besides, PET and Raman spectroscopy also require TDCs with an acceptable resolution of 50 ps [8], [9], [25], [26]. Similar requirements make it possible to design general-purpose TDCs for these applications.

Interpolations [27]–[29] in TDCs use the elements' delay to obtain a higher resolution. The tapped delay line TDC (TDL-TDC) is the mainstream for FPGA-TDCs because cascaded carry-chains are available in modern FPGAs, for example, CARRY4 modules in Xilinx 6-series and 7-series FPGAs [30] and CARRY8 modules in UltraScale and UltraScale+ FPGAs [31]. The TDL-TDC's resolution is determined by the CARRY element's propagation delay and can achieve 10 ps or better [32], [33] in 7-series FPGAs and 5 ps or better [34], [35] in UltraScale FPGAs. Due to the slight propagation delay of CARRY elements, a TDL often requires more than 200 CARRY elements to cover a sampling period. For example, it

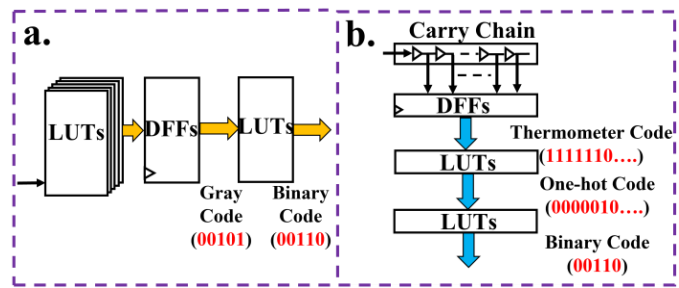

Fig.1. The coding comparison between the (a) GCO-TDC and (b) TDL-TDC.

The research has been supported by the Engineering and Physical Sciences Research Council under EPSRC Grant: EP/L01596X/1, the Royal Society of Edinburgh, and the China Scholarship Council. We would also like acknowledge the support from the Xilinx for donating FPGA develop kits to the research group.

Y. Wang, W. Xie and D. D.-U. Li are with the Faculty of Engineering, University of Strathclyde, Glasgow, G4 0RE, U.K., (e-mail: y.wang.100@strath.ac.uk; wujun.xie@strath.ac.uk; David.li@strath.ac.uk). H. Chen is with the Fraunhofer UK Research Ltd (e-mail: Haochang.Chen@fraunhofer.co.uk).

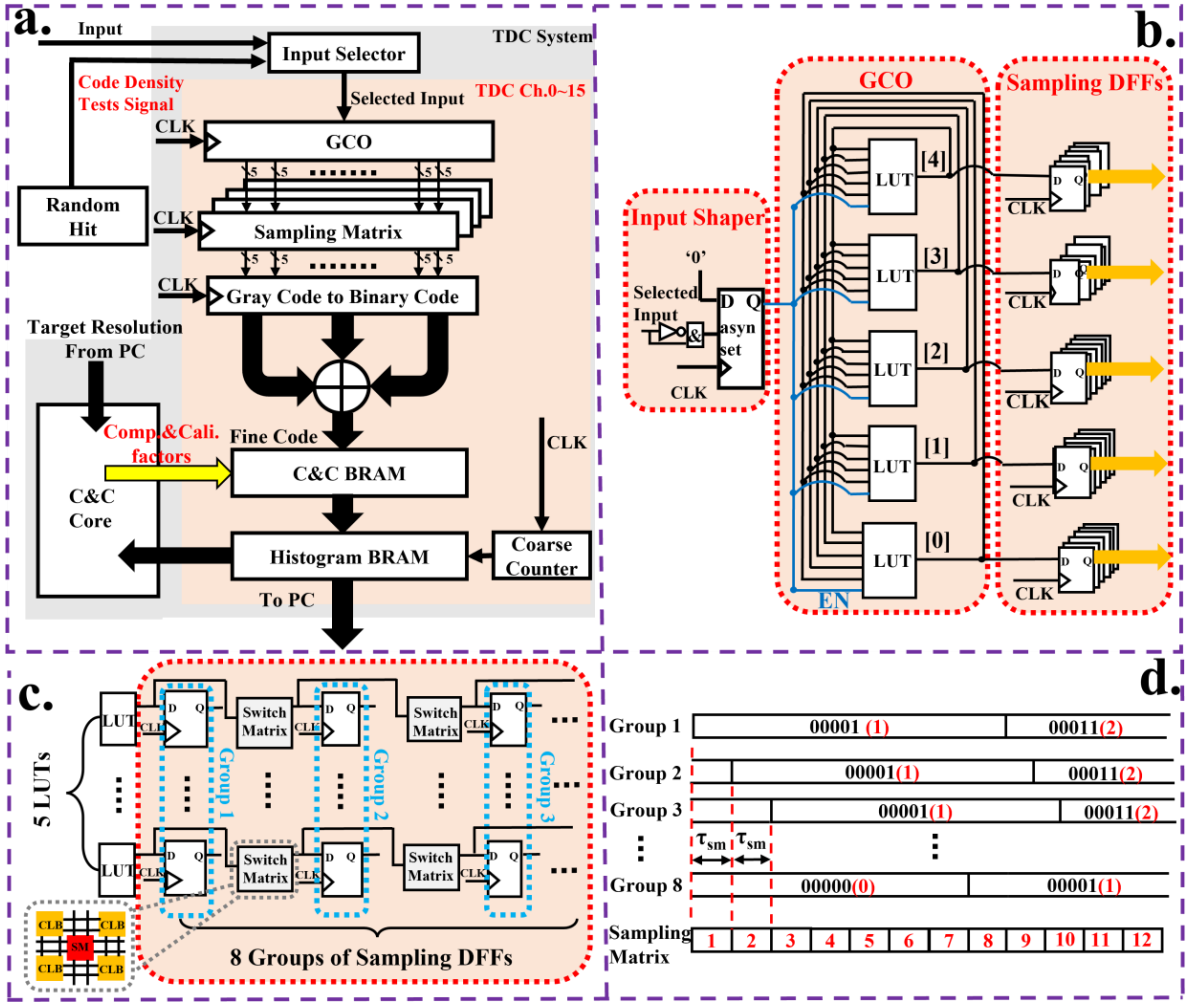

Fig.2. (a) The block diagram of the proposed TDC system. (b) The block diagram of the GCO-TDC (c) The block diagram of the sampling matrix. (d) The concept of the sampling matrix.

requires 50 CARRY4s (200 CARRY elements) in the 7-series FPGA (with a 710MHz sampling clock) [36] and 74 CARRY8s (592 CARRY elements) in the UltraScale FPGA (with a 500MHz sampling clock) [23]. When implementing multichannel designs, the consumption of CARRY elements would be significant (31.26% of CARRY8s are used for 128 channels in Ref. [23]). Besides, in highly-integrated systems, such as LiDAR applications [37], [38], multichannel TDCs are integrated with processing modules in FPGAs. Hence, as a basic unit for arithmetic operations, CARRY elements should be utilized efficiently rather than primarily as delay elements.

Recently, Wu and Xu [39] proposed a gray code oscillator (GCO)-TDC. Compared with TDL-TDCs, this design uses look-up tables (LUTs) rather than carry chains as delay elements. Through outputting gray codes directly, the GCO-TDC avoid tedious coding that uses many logic resources in TDL-TDCs (the comparison is shown in Fig.1). Hence, the GCO-TDC has high hardware utilization efficiency and uses only eight LUTs and eight D-type flip-flops (DFFs) to build one TDC [39]. However, this GCO-TDC's resolution (256 ps) can be further improved, and it has a 1.25 LSB peak-to-peak differential nonlinearity ($DNL_{pk-pk}$). For better linearity, Machado et al. [40] adopted manual routing for the GCO-TDC and improved $DNL_{pk-pk}$ to 0.76 LSB, but the resolution (380.9 ps) significantly increased. To enhance the resolution, Araújo et al. [41] proposed a double-sampling GCO-TDC. This work utilized the double-sampling method to improve the resolution to 69 ps. However, the linearity deteriorates to 1.76 LSB $DNL_{pk-pk}$.

Besides, aforementioned GCO-TDCs [39]–[41] and most previously reported FPGA-TDCs [32], [33], [42] are designed with fixed resolutions. It is not friendly if application requirements change. For broader applications, TDCs in Ref. [23], [43], [44] offer flexible resolutions. However, they all need manual configuration channel-by-channel and chip-by-chip when resolution requirements change. This process is time-consumption and difficult for users unfamiliar with FPGA-TDCs. TDCs in Ref. [45], [46] achieve automatic calibration. But their resolutions are fixed. Hence, an automatic calibration TDC with flexible resolutions is desirable for general applications.

Based on the GCO structure, we proposed a high linearity multichannel TDC with configurable resolutions and automatic calibration. Although we aim for a resolution of 20~100 ps in this paper, it can be extended if needed. The main contributions of this work are:

1) We propose a new sampling matrix structure and dramatically improve the GCO-TDC's resolution.

2) We propose a virtual bin calibration method (VBCM) for online resolution configuration and automatic calibration.
3) We implemented the VBCM by the hardware description language (HDL). Through multiplexing critical components, this core is hardware-efficient.
4) To show our methods, we developed and evaluated 16-channel TDCs in 16nm UltraScale+ XCZU7EV, 20nm UltraScale XCKU040 and 28nm Virtex-7 XC7V690T FPGAs.

This article is structured as follows: Section II describes the architecture and design of the proposed TDC. Section III presents the experimental results, Section IV compares with other designs, and Section V summarizes our TDC.

## II. ARCHITECTURE AND DESIGN

The architecture of the proposed TDC is shown in Fig. 2a. The GCO is the cornerstone of the proposed TDC, and it is responsible for measuring TIs with a coarse counter. Each channel also contains a sampling matrix, a gray-code-to-binary-code converter, a compensation and calibration BRAM (C&C BRAM in Fig. 2a), and a histogram BRAM. Besides, the proposed TDC system contains a compensation and calibration core (C&C Core in Fig. 2a) for calculating compensation and calibration factors (comp.&cali. factors in Fig. 2a) for all 16 channels. The C&C core only works after system launching and configuring the resolution. After calculating, the C&C core loads comp.&cali. factors into the C&C BRAM. With the C&C core and the C&C BRAM, the proposed TDC achieves multi-resolution and high linearity without manual intervention.

### A. GCO-TDC

The GCO is shown in Fig. 2b. Unlike the binary code, only one bit experiences transition between two contiguous states in the gray code. So, the GCO is robust against the "race and competition" phenomenon and can be structured with combinational logic resources [39]. In our design, the GCO is implemented with LUTs. In 7-series and more advanced Xilinx FPGAs, each LUT has up to 6 inputs [30], [31]. One of these inputs is connected to "EN" (highlighted in blue in Fig. 2b) to start and reset the GCO, and the other five inputs are used to get feedback from LUTs' outputs. By instantiating LUTs and DFFs with Vivado primitives [30], [31], five-bit gray codes are generated from five LUTs and then sampled by DFFs.

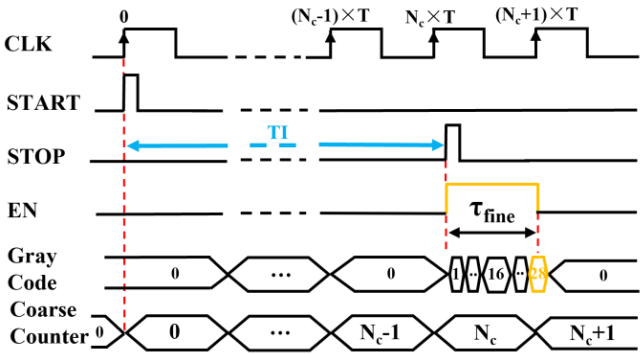

Fig. 3. The timing diagram of the GCO-TDC.

According to GCO's output, the TI between the GCO launching and DFFs sampling can be evaluated. Hence, the GCO can measure the TI even if it is less than one clock period.

To reset the GCO after sampling and work with a coarse counter, as shown in Fig. 2b, an "Input Shaper" is designed to generate the control signal for the GCO. The timing diagram of the proposed GCO-TDC is shown in Fig. 3. In the proposed system, the "START" is synchronous with the sampling clock (CLK in Fig. 3), and the "STOP" is asynchronous with it. When a rising edge of "STOP" comes ("STOP" is the input signal for the "Input Shaper"), the signal "EN" changes to "1"(high-logic level) and keeps this state until the rising edge of the sampling clock. Meanwhile, the GCO launches, and then it resets after being sampled by DFFs. With every rising edge of "STOP", a fine code from the GCO and a coarse code from the coarse counter can be latched. The proposed GCO-TDC can easily extend the measurement range without increasing gray code bits by combining fine and coarse codes. The TI (highlighted in blue in Fig. 3) can be calculated as $TI = (N_c+1) \times T - \tau_{fine}$, where $N_C$ is the coarse code and $\tau_{fine}$ is the time interval corresponding to the fine code.

### B. Sampling Matrix

In the plain GCO-TDC [39], each LUT is sampled by a single DFF. Although this structure is hardware-efficient, it can only deliver a lower resolution. Therefore, a sampling matrix structure (shown in Fig. 2c) is proposed to improve the resolution.

In Fig. 2c, each LUT is sampled by eight DFFs. With eight groups of DFFs, the TDC conducts eight measurements for the same TI in one clock period. Figure 2d shows this method's concept, where $\tau_{sm}$ is the delay of the switch matrix. In plain GCO-TDCs, the resolution is defined as $Q = T/n$. For the GCO-TDC with a sampling matrix, each plain bin is sub-divided with $\tau_{sm}$. Hence, the resolution with a sampling matrix is defined as:

$$LSB_{sm} = \frac{T}{n} \times \frac{1}{M} = \frac{Q}{M}, \quad (1)$$

where $M$ is the amount of DFF groups.

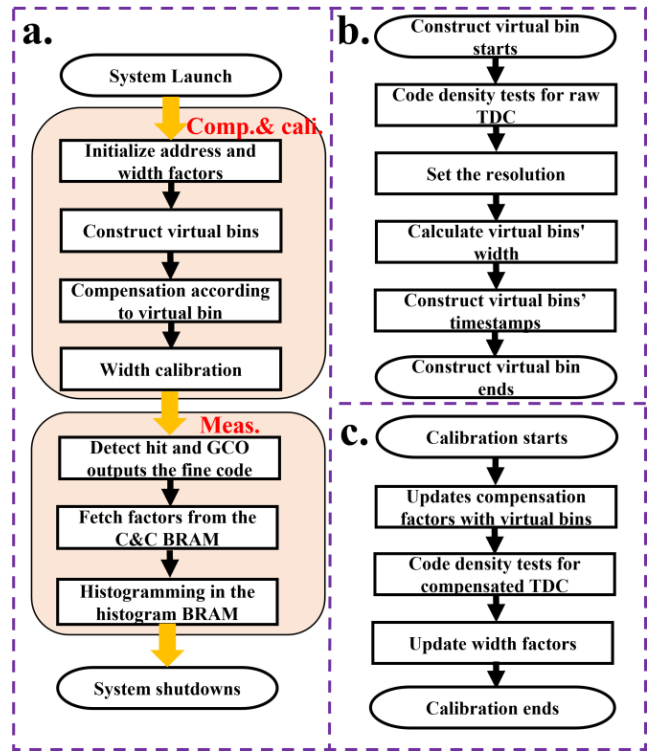

Fig. 4. (a) The workflow the proposed TDC system. (b) The workflow of the virtual bin construction. (c) The workflow of compensation and calibration.

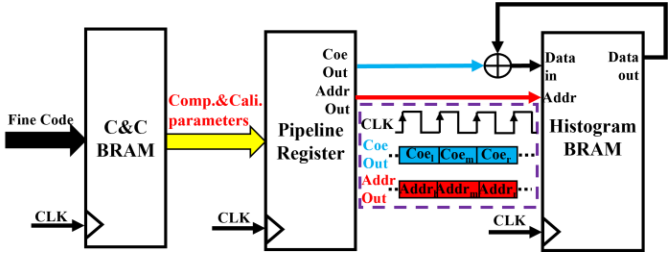

Fig. 5. Hardware implementation of the real-time histogram with the VBCM.

TABLE I compares raw TDCs' resolutions and hardware consumption (without the C&C core) with different sampling factors ($M$) in all three FPGAs. According to the utilization percentage, LUTs' consumption is dominant in hardware consumption. So, we use the consumption of LUTs to evaluate hardware consumption in Eq (2). To find a balance between the resolution and hardware consumption when we increase $M$, the normalized efficiency of resolution improvement is proposed and calculated as:

$$E_M = \frac{LSB_{M[i-1]} - LSB_{M[i]}}{LSB_{plain}} \Big/ \frac{LUT_{M[i]} - LUT_{M[i-1]}}{LUT_{plain}}, M=2^i \quad (2)$$

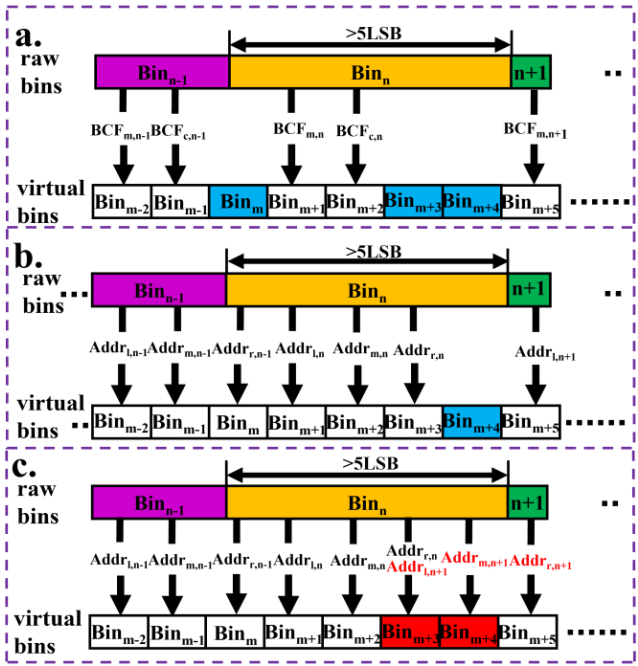

Fig. 6. Compensation in (a) the mixed calibration, (b) the weighted histogram calibration and (c) the virtual bin calibration methods.

TABLE I
COMPARISONS OF PERFORMANCE AND HARDWARE UTILIZATION WITH DIFFERENT SAMPLING FACTORS ($M$)

| $M$ | $E_M$ | LSB (ps) | Used LUTs (%)[1] | Used DFFS (%)[2] |
|---|---|---|---|---|
| *UltraScale+* | | | | |
| 1 | N/A[3] | 158.03 | 117 (0.05%) | 177 (0.04%) |
| 2 | 2.56 | 79.01 | 139 (0.06%) | 194 (0.04%) |
| 4 | 0.96 | 40.23 | 170 (0.07%) | 221 (0.05%) |
| **8** | **0.31** | **19.41** | **222 (0.10%)** | **268 (0.06%)** |
| 16 | 0.08 | 10.51 | 295 (0.13%) | 354 (0.08%) |
| *UltraScale* | | | | |
| 1 | N/A[3] | 267.09 | 146 (0.06%) | 233 (0.05%) |
| 2 | 6.50 | 136.39 | 157 (0.06%) | 238 (0.05%) |
| 4 | 1.76 | 68.93 | 178 (0.07%) | 248 (0.05%) |
| **8** | **0.44** | **35.42** | **220 (0.09%)** | **268 (0.06%)** |
| 16 | 0.15 | 19.02 | 279 (0.12%) | 380 (0.08%) |
| *Virtex-7* | | | | |
| 1 | N/A[3] | 256.41 | 146 (0.03%) | 233 (0.03%) |
| 2 | 5.73 | 125.69 | 159 (0.04%) | 238 (0.03%) |
| 4 | 11.69 | 64.10 | 162 (0.04%) | 248 (0.03%) |
| **8** | **0.43** | **32.54** | **204 (0.05%)** | **268 (0.03%)** |
| 16 | 0.11 | 17.05 | 281 (0.06%) | 380 (0.04%) |

[1] Percentage of LUTs' utilization; [2] Percentage of DFFs' utilization; [3] Not available.

where $LSB_M$ and $LUT_M$ are, respectively, the TDC's resolution and consumption of LUTs with an $M$-order sampling matrix, and $LSB_{plain}$ and $LUT_{plain}$ are the resolution and LUTs' consumption of the plain GCO-TDC ($M = 1$). When $M$ increases, $E_M$ decreases in UltraScale and UltraScale+ FPGAs. However, it shows a different trend in the Virtex-7 FPGA. $E_M$ in the Virtex-7 FPGA reaches its peak value when $M = 4$ due to a different FPGA architecture. Overall, $E_M$ is close to 0 when $M = 16$ in all three FPGAs, meaning the efficiency of resolution improvement is low when $M$ increases from 8 to 16. Hence, we choose $M = 8$, considering the trade-off between performance and hardware consumption.

### C. Virtual Bin Calibration Method

Our previous work [46] presented a PS-based architecture to achieve automatic calibration. However, this design is device-dependent, and its resolution is fixed. Here we propose a virtual bin calibration method for automatic calibration and online resolution configuration in common FPGA devices. For this work, we will demonstrate the proposed method in three different FPGA devices manufactured in 16nm, 20nm, and 28nm CMOS processes.

The workflow of the VBCM is shown in Fig. 4a. After the system starts and configures the resolution, the C&C core calculates compensation factors ($Addr_l$, $Addr_m$ and $Addr_r$) and width calibration factors ($Coe_l$, $Coe_m$ and $Coe_r$). Then these factors are loaded to the C&C BRAM in the compensation and calibration stage (Comp.&cali. in Fig. 4a). In the measurement stage (Meas. in Fig. 4a), indexed by a fine code, factors are delivered from the C&C BRAM to the histogram BRAM. During this procedure, width calibration factors ($Coe_l$, $Coe_m$ and $Coe_r$) are seriatim added to the corresponding bins of the histogram BRAM indexed by the compensation factors ($Addr_l$, $Addr_m$ and $Addr_r$) shown in Fig.5. Through this, real-time calibration and resolution configuration can be achieved.

Calculations of compensation and width calibration factors contain: 1) construction of virtual bins and 2) calculations according to virtual bins. These two steps are both based on code density tests. Figure 4b shows the workflow of virtual bins

```
start_point=max[Addr_l[k-1], Addr_m[k-1], Addr_r[k-1]];
if (T_raw[k] ≤ T_vir[start_point] )
    Addr_l[k]= start_point ;
elseif (T_raw[k] ≤ T_vir[start_point+1] )
    Addr_l[k]= start_point ;
    Addr_m[k]= start_point+1 ;
elseif (T_raw[k] ≤ T_vir[start_point+2] )
    Addr_l[k]= start_point ;
    Addr_m[k]= start_point+1 ;
    Addr_r[k]= start_point+2 ;
else
    Addr_l[k]= start_point+1 ;
    Addr_m[k]= start_point+2 ;
    Addr_r[k]= start_point+3 ;
```

Fig. 7. The pseudo-codes of compensation in the virtual bin calibration method.

construction. With the target resolution $R_{conf}$, the number of hits collected at a virtual bin should be:

$$hit_{vir} = \frac{R_{conf}}{T} \times \tilde{N} = \frac{\tilde{N}}{n_{vir}}, \quad (3)$$

where $n_{vir}$ ($n_{vir} \leq n$) is the number of virtual bins in a sampling period and $\tilde{N}$ is the number of random hits for code density tests. Hence, the "timestamp" of the *m*-th virtual bin (number of hits collected until the *m*-th virtual bin) and the "timestamp" of the *k*-th raw bin can be defined, respectively, as:

$$T_{vir}[m] = hit_{vir} \times m, \quad m \in [1, n_{vir}], \quad (4)$$
$$T_{raw}[k] = \sum_1^k hit_{raw}[j], \; k \in [1,n], \quad (5)$$

where $hit_{raw}[j]$ is the number of hits at the *j*-th raw bin.

With $T_{vir}$ and $T_{raw}$, similar to the mixed calibration [35] and weighted calibration [46] methods, the compensation factors can be calculated. However, both the mixed calibration [35] and weighted calibration [46] methods have a limited compensation range, causing "missing bins" (highlighted in blue in Fig. 6a and Fig. 6b). Hence, this work proposes a new missing-bin-free compensation strategy. With three compensation factors (Addr_l, Addr_m and Addr_r), the concept of the proposed compensation strategy is shown in Fig. 6c. In most cases, ultra-narrow bins (< 1 LSB, highlighted in green in Fig. 6c) and regular bins (1 ~ 2 LSB, highlighted in purple in Fig. 6c) could neighbor ultra-wide bins (highlighted in yellow in Fig. 6c). However, earlier studies [35], [46] did not utilize all compensation factors of ultra-narrow bins (only BCF_{m,n+1} is used in Fig. 6a and only Addr_{l,n+1} is used in Fig. 6b). To utilize "idle" compensation factors, in Fig. 6c, Addr_{l,n+1} and Addr_{m,n+1} are used to remap to virtual bins covered by Bin_n (Bin_{m+3} and Bin_{m+4}, highlighted in red in Fig. 6c). The pseudo-codes for compensation factor calculations are shown in Fig. 7. After updating compensation factors according to virtual bins, the width calibration factors can be calculated in Fig. 4c. Code density tests are conducted again for the compensated TDC. Then, with the number of hits collected at each bin, width calibration factors can be calculated as:

$$Coe_{l,m,r}[k] = \frac{\tilde{N}}{n_{vir}} \times \frac{1}{hit_{com}[i]} \quad i = Addr_{l,m,r}[k], \quad (6)$$

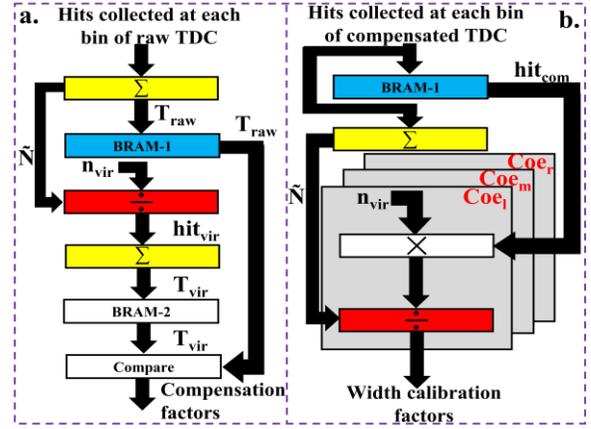

Fig. 8. Hardware implementation of (a) compensation factor calculations, and (b) width calibration factor calculations.

where $hit_{com}[i]$ is the number of hits collected at the *i*-th bin of the compensated TDC.

### D. Hardware implementation of the VBCM

The hardware implementation of the VBCM contains: 1) the implementation of the real-time histogram and 2) the implementation of the C&C core. The implementation of the real-time histogram is shown in Fig. 5. For low BRAM-consumption, a pipeline structure is utilized. In the C&C BRAM, three groups of factors are merged and stored in an address to reduce hardware consumption. With pipeline registers, merged factors are separated and delivered to the histogram BRAM within three system clock periods. Compared with consuming two histogram BRAMs in the mixed calibration method [35] and three histogram BRAMs in our previous work [46], only one histogram BRAM is required in this design.

The C&C core is responsible for calculating compensation and width calibration factors. It is composed of two modules: 1) a compensation factor calculation (CFC) module and 2) a width calibration factor calculation (WCFC) module. The WCFC module works after the CFC module. Hence, we multiplex some components in these two modules to achieve low resource consumption. The hardware implementation of the CFC module is shown in Fig. 8a. After code density tests, $T_{vir}$ and $T_{raw}$ are calculated according to Eqs (4) and (5) and stored in BRAM-2 and BRAM-1, respectively. When all $T_{vir}$ and $T_{raw}$ are calculated, they are output from respective BRAMs and calculated according to pseudo-codes shown in Fig.7 to obtain compensation factors (Addr_l, Addr_m and Addr_r). After compensation is complete, compensation factors are updated into the C&C BRAM. Then code density tests are conducted again for the compensated TDC, and results are stored in BRAM-1, as shown in Fig. 8b. With the bin widths of the compensated TDC, the width calibration factors ($Coe_l$, $Coe_m$ and $Coe_r$) can be calculated according to Eq (6). In the WCFC module, we reuse BRAM-1, the accumulator and the divider (highlighted in blue, yellow and red in Fig.7) previously used in the CFC module. Moreover, three width calibration factors are calculated similarly. Therefore, the multiplier-divider component (highlighted in gray) is also multiplexed in the WCFC module.

TABLE II
RESULTS OF THE VBCM WITH A DIFFERENT $\bar{M}$

| 16nm UltraScale+ | | | | | | |
|---|---|---|---|---|---|---|
| $\bar{M}$ | 1 | 2 | 3 | 4 | 5 | 6 |
| DNL_{pk-pk} (LSB) | 0.68 | 0.39 | 0.20 | 0.12 | 0.10 | 0.08 |
| INL_{pk-pk} (LSB) | 3.21 | 1.39 | 0.50 | 0.42 | 0.49 | 0.31 |
| 20nm UltraScale | | | | | | |
| $\bar{M}$ | 1 | 2 | 3 | 4 | 5 | 6 |
| DNL_{pk-pk} (LSB) | 0.68 | 0.38 | 0.20 | 0.16 | 0.12 | 0.12 |
| INL_{pk-pk} (LSB) | 2.56 | 0.95 | 0.77 | 0.48 | 0.36 | 0.35 |
| 28nm Virtex-7 | | | | | | |
| $\bar{M}$ | 1 | 2 | 3 | 4 | 5 | 6 |
| DNL_{pk-pk} (LSB) | 0.65 | 0.36 | 0.20 | 0.16 | 0.09 | 0.10 |
| INL_{pk-pk} (LSB) | 3.11 | 1.35 | 0.73 | 0.62 | 0.26 | 0.34 |

TABLE III
LINEARITY OF THE PROPOSED TDC WITH DIFFERENT RESOLUTIONS

| | | $LSB$ (ps) | $DNL_{pk\text{-}pk}$ (LSB) | $\sigma_{DNL}$ (LSB) | $INL_{pk\text{-}pk}$ (LSB) | $\sigma_{INL}$ (LSB) | $\omega_{eq}$ (ps) | $\sigma_{eq}$ (LSB) |
|---|---|---|---|---|---|---|---|---|
| UltraScale+ 16nm | Raw-TDC | 19.41 ($n = 228$) | 3.98 | 0.96 | 7.17 | 1.39 | 43.11 | 0.64 |
| | VBCM-TDC | 20.97 ($n_{vir} = 211$) | 0.09 | 0.01 | 0.20 | 0.04 | 20.98 | 0.29 |
| | | 29.90 ($n_{vir} = 148$) | 0.05 | 0.01 | 0.12 | 0.02 | 29.91 | 0.29 |
| | | 39.86 ($n_{vir} = 111$) | 0.05 | 0.01 | 0.12 | 0.03 | 39.86 | 0.29 |
| | | 50.28 ($n_{vir} = 88$) | 0.05 | 0.01 | 0.11 | 0.02 | 50.29 | 0.29 |
| | | 80.45 ($n_{vir} = 55$) | 0.03 | 0.01 | 0.09 | 0.02 | 80.46 | 0.29 |
| UltraScale 20nm | Raw-TDC | 35.42 ($n = 181$) | 4.76 | 0.93 | 12.38 | 2.92 | 76.94 | 4.72 |
| | VBCM-TDC | 36.01 ($n_{vir} = 178$) | 0.08 | 0.01 | 0.14 | 0.03 | 36.02 | 0.29 |
| | | 40.06 ($n_{vir} = 160$) | 0.07 | 0.01 | 0.12 | 0.03 | 40.07 | 0.29 |
| | | 50.08 ($n_{vir} = 128$) | 0.05 | 0.01 | 0.13 | 0.03 | 50.09 | 1.00 |
| | | 80.13 ($n_{vir} = 80$) | 0.05 | 0.01 | 0.08 | 0.02 | 80.15 | 0.29 |
| | | 100.16 ($n_{vir} = 64$) | 0.04 | 0.01 | 0.11 | 0.03 | 100.18 | 0.29 |
| Virtex-7 28nm | Raw-TDC | 32.54 ($n = 197$) | 5.48 | 0.95 | 11.23 | 2.35 | 72.96 | 0.65 |
| | VBCM-TDC | 34.84 ($n_{vir} = 184$) | 0.07 | 0.02 | 0.29 | 0.08 | 34.85 | 0.29 |
| | | 40.06 ($n_{vir} = 160$) | 0.07 | 0.01 | 0.18 | 0.04 | 40.07 | 0.29 |
| | | 50.08 ($n_{vir} = 128$) | 0.07 | 0.01 | 0.16 | 0.04 | 50.09 | 0.29 |
| | | 80.13 ($n_{vir} = 80$) | 0.04 | 0.01 | 0.15 | 0.03 | 80.15 | 0.29 |
| | | 100.16 ($n_{vir} = 64$) | 0.05 | 0.01 | 0.11 | 0.03 | 100.18 | 0.29 |

All data is presented and calculated in the fixed-point format to reduce hardware consumption in our design. Considering errors when the decimal parts are discarded, we use the last $\bar{M}$ bits of $Coe$ to present decimals. Then, calibration factors can be calculated as:

$$\overline{Coe_{l,m,r}[k]} = Coe_{l,m,r}[k] \times 2^{\bar{M}} = \frac{\tilde{N}}{n_{vir}} \times \frac{2^{\bar{M}}}{hit_{com}[i]}, i = Addr_{l,m,r}[k]. \quad (7)$$

For selecting an appropriate $\bar{M}$, we used MATLAB to simulate the VBCM with a different $\bar{M}$ (setting $n_{vir} = n$). Results in TABLE II indicate that when $\bar{M}$ increases from 5 to 6, the linearity cannot be further improved in UltraScale and Virtex-7 FPGAs, but only peak-to-peak INL (INL$_{pk\text{-}pk}$) is enhanced in the UltraScale+ FPGA. Hence, we choose $\bar{M} = 5$ for all three FPGAs for an optimized design.

## III. EXPERIMENTAL RESULTS

We implemented the proposed TDC in ZCU104 [47] (16nm UltraScale+), KCU105 [48] (20nm UltraScale) and NetFPGA SUME [49] (28nm Virtex-7) evaluation boards, respectively. Random hits for code density tests were generated by an SRS CG-635 (Stanford Research Systems), and TDCs' clocks are from low-jitter crystal oscillators on the boards (IDT-8T49 in ZCU104, SI-570 in KCU105 and DSC-1103 in NetFPGA SUME). The frequencies of TDCs' clocks are 226MHz (ZCU104), 156MHz (KCU105) and 156MHz (NetFPGA SUME), respectively, due to different GCOs' oscillation frequencies in three FPGAs. The temperature and voltage were maintained in experiments.

### A. Resolution configuration and linearity

We evaluated our configurable TDCs with different resolution options. The linearity is characterized by DNL, INL and their standard deviations ($\sigma_{DNL}$ and $\sigma_{INL}$). Besides, Wu [50] also proposed the equivalent bin-width ($\omega_{eq}$) and its deviation ($\sigma_{eq}$) to evaluate the TDC's linearity. They are:

$$\sigma_{eq}^2 = \sum_{i=1}^{n} \left( \frac{W[i]^2}{12} \times \frac{W[i]}{W_{total}} \right), \quad (8)$$

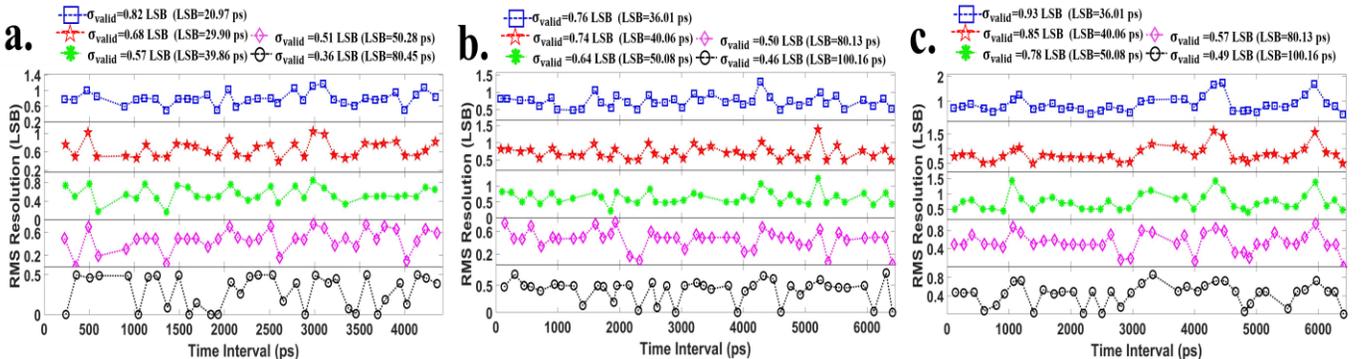

Fig.9. RMS resolutions with different resolution configurations in (a) 16nm UltraScale+, (b) 20nm UltraScale, and (c) 28nm Virtex-7 FPGAs.

$$\omega_{eq} = \sigma_{eq} \times \sqrt{12} = \sqrt{\sum_{i=1}^{n} \frac{W[i]^3}{W_{total}}}, \quad (9)$$

where $W_{total} = \sum_{i=1}^{n} W[i]$. The experimental results are summarized in TABLE III. With the VBCM, the proposed TDC's linearity is improved significantly. The 16nm UltraScale+ TDC has DNL$_{pk-pk}$ enhanced by more than 44-fold (from 3.98 LSB to less than 0.09 LSB), and INL$_{pk-pk}$ enhanced by more than 35-fold (from 7.17 LSB to less than 0.20 LSB). Besides, we have also achieved significant improvements in 20nm UltraScale and 28nm Virtex-7 FPGAs, with DNL$_{pk-pk}$ respectively improved by more than 59-fold (from 4.76 LSB to less than 0.08 LSB) and 78-fold (from 5.48 LSB to less than 0.07 LSB), and INL$_{pk-pk}$ respectively improved by more than 88-fold (from 12.38 LSB to less than 0.14 LSB) and 38-fold (from 11.23 LSB to less than 0.29 LSB). Results indicate that the proposed TDC has high linearity in different resolutions.

*B. Time Interval Tests*

The RMS resolution can be evaluated by the standard deviation ($\sigma$) of measurements for the same TI. It is defined as $\sigma^2 = \sum_{i=1}^{N_T} \frac{(x_i-\mu)^2}{N_T-1}$, where $x_i$ is the $i$-th output and $\mu$ is the averaged value for $N_T$ measurements when the TI is fixed.

To avoid jitter introduced by input signals, we utilized programmable input delay elements inside FPGAs (IDELAY3 in UltraScale+ and UltraScale FPGAs, and IDELAY2 in the Virtex-7 FPGA) to generate controllable delays.

The RMS resolutions for different resolution configurations are shown in Fig. 9. We conducted time interval tests with different intervals (less than one system clock period). However, for each resolution option with different intervals, the averaged RMS resolution or the maximum RMS resolution cannot represent the RMS resolution because they overestimate or underestimate it [23]. Hence, the valid RMS resolution ($\sigma_{valid}$) is used, defined as $\sigma_{valid}^2 = \sum_{1}^{H} \frac{\sigma_i^2}{H}$ [23], where $\sigma_i$ is the standard deviation of measurements for the $i$-th fixed time interval and $H$ is the number of different time intervals. With the resolution improved, the valid RMS resolution deteriorates in all three FPGAs. The UltraScale+ version achieves the best valid RMS resolution of 0.36 LSB when LSB=80.45 ps. The UltraScale and Virtex-7 versions, respectively, can achieve valid RMS resolutions of 0.46 LSB and 0.49 LSB when LSB=100.16 ps.

*C. Multichannel Design*

We implemented the proposed TDCs in all three FPGAs. In each channel, an 18k-BRAM is used as the histogram BRAM, and a 36k-BRAM is used as the C&C BRAM. Besides, less than 230 LUTs and 270 DFFs are required to build the GCO-TDC with a sampling matrix. For the C&C core, two 18k-BRAMs, no more than 3800 LUTs and 1600 DFFs are used to calculate comp.&cali. factors. The resources required for the proposed

TABLE IV
HARDWARE RESOURCES UTILIZATION

|  |  | LUT | DFF | CARRY[1] | BRAM[2] |
|---|---|---|---|---|---|
| **UltraScale+ 16 nm** | Available | 230400 | 460800 | 28800 | 312 |
|  | 1-ch. | 222 | 268 | 9 | 1.5 |
|  | 16-ch. | 3548 | 4288 | 144 | 24 |
|  | C&C core | 3719 | 1596 | 310 | 1 |
| **UltraScale 20 nm** | Available | 242400 | 484800 | 30300 | 600 |
|  | 1-ch. | 220 | 268 | 9 | 1.5 |
|  | 16-ch. | 3518 | 4288 | 144 | 24 |
|  | C&C core | 3725 | 1569 | 310 | 1 |
| **Virtex-7 28 nm** | Available | 433200 | 866400 | 108300 | 1470 |
|  | 1-ch. | 204 | 268 | 18 | 1.5 |
|  | 16-ch. | 3263 | 4288 | 288 | 24 |
|  | C&C core | 3753 | 1597 | 574 | 1 |

[1] CARRY8s in UltraScale+ and UltraScale FPGAs, and CARRY4s in Virtex-7 FPGA; [2] 36K-BRAM.

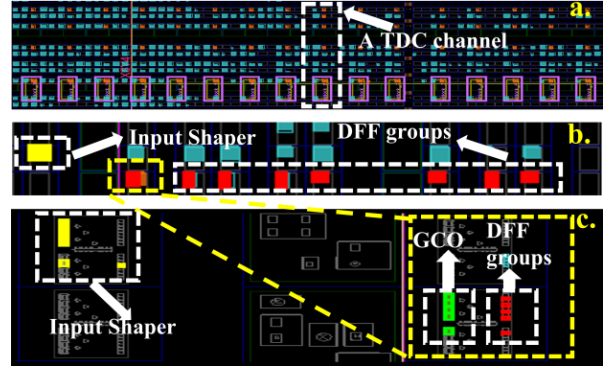

Fig. 10. Implementation layouts of (a) 16 channels, (b) a single channel, and (c) an input shaper and a GCO.

TABLE V
DNL$_{PK-PK}$ OF 16-CHANNEL TDCS IN 16NM, 20NM, AND 28NM FPGAS

| LSB (ps) | 0 | 1 | 2 | 3 | 4 | 5 | 6 | 7 | 8 | 9 | 10 | 11 | 12 | 13 | 14 | 15 | Ave. |
|---|---|---|---|---|---|---|---|---|---|---|---|---|---|---|---|---|---|
|  | LSB |  |  |  |  |  |  |  |  |  |  |  |  |  |  |  |  |
| *UltraScale+* |  |  |  |  |  |  |  |  |  |  |  |  |  |  |  |  |  |
| 20.97 | 0.09 | 0.07 | 0.10 | 0.08 | 0.07 | 0.18 | 0.08 | 0.07 | 0.11 | 0.06 | 0.07 | 0.07 | 0.07 | 0.13 | 0.07 | 0.08 | 0.09 |
| 29.90 | 0.05 | 0.06 | 0.06 | 0.06 | 0.05 | 0.08 | 0.06 | 0.07 | 0.06 | 0.06 | 0.06 | 0.07 | 0.06 | 0.06 | 0.06 | 0.05 | 0.06 |
| 39.86 | 0.05 | 0.04 | 0.05 | 0.04 | 0.05 | 0.05 | 0.06 | 0.05 | 0.05 | 0.06 | 0.05 | 0.05 | 0.04 | 0.05 | 0.05 | 0.06 | 0.05 |
| 50.28 | 0.05 | 0.04 | 0.04 | 0.04 | 0.05 | 0.05 | 0.04 | 0.04 | 0.04 | 0.04 | 0.04 | 0.05 | 0.05 | 0.05 | 0.04 | 0.05 | 0.04 |
| 80.45 | 0.03 | 0.04 | 0.04 | 0.04 | 0.04 | 0.04 | 0.04 | 0.04 | 0.04 | 0.04 | 0.04 | 0.04 | 0.04 | 0.04 | 0.04 | 0.04 | 0.04 |
| *UltraScale* |  |  |  |  |  |  |  |  |  |  |  |  |  |  |  |  |  |
| 36.01 | 0.08 | 0.16 | 0.08 | 0.13 | 0.08 | 0.08 | 0.08 | 0.09 | 0.07 | 0.18 | 0.09 | 0.07 | 0.08 | 0.08 | 0.21 | 0.09 | 0.10 |
| 40.06 | 0.07 | 0.11 | 0.10 | 0.20 | 0.10 | 0.02 | 0.06 | 0.12 | 0.09 | 0.10 | 0.07 | 0.09 | 0.09 | 0.10 | 0.25 | 0.16 | 0.11 |
| 50.08 | 0.05 | 0.15 | 0.14 | 0.05 | 0.17 | 0.07 | 0.06 | 0.05 | 0.07 | 0.06 | 0.07 | 0.09 | 0.06 | 0.09 | 0.06 | 0.05 | 0.08 |
| 80.13 | 0.05 | 0.06 | 0.05 | 0.08 | 0.05 | 0.15 | 0.05 | 0.05 | 0.05 | 0.05 | 0.05 | 0.05 | 0.09 | 0.06 | 0.05 | 0.06 | 0.06 |
| 100.16 | 0.04 | 0.18 | 0.05 | 0.06 | 0.04 | 0.16 | 0.05 | 0.04 | 0.05 | 0.05 | 0.05 | 0.04 | 0.06 | 0.04 | 0.05 | 0.05 | 0.06 |
| *Virtex-7* |  |  |  |  |  |  |  |  |  |  |  |  |  |  |  |  |  |
| 34.84 | 0.07 | 0.07 | 0.06 | 0.06 | 0.10 | 0.07 | 0.07 | 0.06 | 0.05 | 0.11 | 0.06 | 0.15 | 0.09 | 0.07 | 0.07 | 0.10 | 0.08 |
| 40.06 | 0.07 | 0.11 | 0.09 | 0.07 | 0.09 | 0.08 | 0.08 | 0.08 | 0.09 | 0.15 | 0.08 | 0.08 | 0.07 | 0.09 | 0.06 | 0.09 | 0.09 |
| 50.08 | 0.07 | 0.07 | 0.06 | 0.06 | 0.05 | 0.06 | 0.06 | 0.07 | 0.05 | 0.06 | 0.06 | 0.07 | 0.06 | 0.06 | 0.06 | 0.07 | 0.06 |
| 80.13 | 0.04 | 0.04 | 0.05 | 0.05 | 0.05 | 0.05 | 0.04 | 0.05 | 0.05 | 0.05 | 0.05 | 0.05 | 0.05 | 0.05 | 0.05 | 0.05 | 0.05 |
| 100.16 | 0.05 | 0.04 | 0.05 | 0.04 | 0.05 | 0.05 | 0.05 | 0.04 | 0.04 | 0.04 | 0.04 | 0.04 | 0.05 | 0.04 | 0.04 | 0.05 | 0.04 |

TABLE VI
COMPARISON OF RECENTLY PUBLICATED FPGA-TDCs

| Ref-year | Methods | Devi. Proc. (nm) | LSB (ps) | $\omega_{eq}$ (ps) | RMS Resol. (ps) | DNL(LSB) | INL(LSB) | LUT (%)[1] | DFF (%)[1] | Carry (%)[1] | 36K-BRAM | Auto/manual Cali. |
|---|---|---|---|---|---|---|---|---|---|---|---|---|
| **TDL-TDCs** | | | | | | | | | | | | |
| [32]-17 | Tuned-TDL, Direct Histogram, Bin-width Cali. | 28 | 10.50 | 10.55 | 4.42 | [-0.04,0.04] | [-0.09,0.04] | N/S[2] | N/S[2] | N/S[2] | N/S[2] | Manual |
| [35]-19 | Tuned-TDL, Sub-TDL, Mixed Calibration | 28 | 10.54 | 10.55 | 14.59 | [-0.05, 0.08] | [-0.09, 0.11] | 1145 0.26 | 1916 0.22 | N/S[2] | 1.5 | Manual |
|  |  | 20 | 5.02 | 5.03 | 7.80 | [-0.12, 0.11] | [-0.18, 0.46] | 703 0.29 | 1195 0.24 | 80[10] 0.26 |  |  |
| [45]-21 | Slide Scale, Gain & Error cal., Moving Ave. | 28 | 4.88 | N/S[2] | 2.90~8.03 | [-0.10, 0.15] | [-0.23, 0.28] | 2962 N/S[2] | 4157 N/S[2] | N/S[1] | N/S[1] | Auto |
| [23]-21 | Mixed-binning | 20 | 51.28 83.33 105.26 | 51.29 83.34 105.26 | 15.89 21.67 26.32 | [-0.018, 0.021] [-0.017, 0.016] [-0.008, 0.008] | [-0.019,0.035] [-0.028,0.003] [-0.009,0.007] | 663 0.27 | 1124 0.23 | 74[10] 0.24 | 2.5 | Manual |
| [46]-21 | Sub-TDL, AC-WU | 28 | 9.83 | 9.85 | 13.86 | [-0.14, 0.16] | [-0.25, 0.42] | 764 1.44 | 1095 1.02 | 50[11] 0.38 | 2 | Auto |
| [44]-21 | Half single-chain, real States-based Coding | 16 | 5 21.56 87.74 | N/S[2] N/S[2] N/S[2] | 19[3] 30.18[3] 105.29[3] | [-0.99, 1.44] [-0.16, 0.19] [-0.07, 0.05] | [-2.84, 1.62] [-0.50, 0.33] [0.00, 0.11] | N/S[2] N/S[2] N/S[2] | N/S[2] N/S[2] N/S[2] | N/S[2] N/S[2] N/S[2] | N/S[2] N/S[2] N/S[2] | N/S[2] |
| **Other TDCs** | | | | | | | | | | | | |
| [42]-20 | Bidirectional, RO Vernier | 65 | 24.50 | N/S[2] | 28.00 | [-0.20,0.25] | [0.03,0.82] | 172 N/S[2] | 986 N/S[2] | N/S[2] | N/S[2] | N/S[2] |
| [43]-21 | NUMMP, Timing Scale Marking | 28 | 1.87 | N/S[2] | 2.79 | [-0.54,1.30] | [-2.21,3.51] | 1679 0.82 | 1103 0.27 | N/S[2] | 12 | N/S[2] |
|  |  |  | 11.24 | N/S[2] | 8.07 | [-0.43,0.26] | [-0.55,-0.30] | 1328 0.65 | 857 0.2 | N/S[2] | 12 |  |
|  |  |  | 20.00 | N/S[2] | 12.81 | [-0.05,0.06] | [-0.15,0.08] | 634 0.41 | 828 0.16 | N/S[2] | 12 |  |
| **GCO-TDCs** | | | | | | | | | | | | |
| [39]-2019 | GCO, Bin-by-bin Cali. | 28 | 256 | N/S[2] | 155 | [-0.53,0.72][6] | N/S[2] | 8 N/S[2] | 8 N/S[2] | N/S[2] | N/S[2] | Manual |
| [40]-2020 | GCO, Manual Routing | 28 | 380.9 | N/S[2] | 290 | [-0.38,0.38] | [0.01,0.70] | 6 N/S[2] | 10 N/S[2] | N/S[2] | N/S[2] | Manual |
| [41]-2021 | GCO, Double Sampling | 16 | 69 | N/S[2] | 54.99 | [-0.95,0.81] | [-1.01,0.49] | 5 N/S[2] | 19 N/S[2] | N/S[2] | N/S[2] | Manual |
| **This TDC** | GCO, Sampling Matrix, VBCM | 16 | 20.97[4] | 20.98 | 17.11[5] | [-0.055, 0.034] 0.087[7] | [-0.196, 0.000] 0.224[7] | 222[8] 0.10 455[9] 0.20 | 268[8] 0.06 368[9] 0.08 | 9[10] 0.03 | 1.5[12] | Auto |
|  |  | 20 | 36.01[4] | 36.02 | 27.37[5] | [-0.036, 0.046] 0.102[7] | [-0.057, 0.081] 0.262[7] | 220[8] 0.09 453[9] 0.19 | 268[8] 0.06 367[9] 0.08 | 9[10] 0.03 | 1.5[12] |  |
|  |  | 28 | 34.84[4] | 34.85 | 32.33[5] | [-0.033, 0.034] 0.078[7] | [-0.016, 0.277] 0.203[7] | 204[8] 0.05 437[9] 0.10 | 268[8] 0.03 368[9] 0.04 | 18[11] 0.02 | 1.5[12] |  |

[1] Percentage of resource utilization for the target device; [2] N/S=not specified; [3] FWHM of the residual; [4] Proposed TDCs' best resolution in this device; [5] Valid RMS resolution; [6] Approximate values from figures presented in literature; [7] Averaged peak-to-peak DNL or INL of multichannel TDCs. [8] Each channel's logic resource consumption without the C&C core; [9] Each channel's average logic resource consumption with the C&C core; [10] CARRY8s; [11] CARRY4s; [12] Each channel's BRAM consumption.

TDCs are summarized in TABLE IV. It indicates our design is more hardware-effective compared with TDL-TDCs presented in Ref. [23], [32] and [46], and has similar logic resource consumption compared with the RO-TDC presented in Ref. [42] (a comparison shown in Table VI).

Implementation layouts in the UltraScale+ FPGA are shown in Fig. 10. Each "Input Shaper" is constrained near the corresponding GCO to minimize jitters introduced by routing resources. Moreover, DFFs are manually placed to enhance linearity. In the UltraScale+ FPGA, DFF groups are placed contiguously (SLICE XnYm and SLICE X(n+1)Ym) in a row. In UltraScale and Virtex-7 FPGAs, DFF groups are placed at a fixed distance (SLICE XnYm and SLICE X(n+2)Ym).

Code density tests were conducted for 16-channel TDCs in three FPGAs. The linearity with different resolutions is summarized in TABLE V, showing the proposed TDC has high linearity and good uniformity.

## IV. COMPARISONS AND DISCUSSIONS

TABLE VI summarizes recently published FPGA-TDCs and the proposed TDC. As shown in Table VI, the TDL-TDC is the mainstream design. Other advanced architectures like the ring-oscillator-based (RO-based) Vernier [42] and the nonuniform monotonic multiphase (NUMMP) architectures [43] were also well-developed. However, we have further developed the new

GCO-TDC architecture (firstly published in Ref. [39] by Wu and Xu).

Unlike high resolution (<10 ps) TDCs aimed for scientific applications, our design aims for multichannel industrial LiDAR applications with variable resolutions and high linearity. Compared with TDL-based and NUMMP-based TDCs for similar specifications (for example, a 20 ps resolution or high linearity), the proposed TDC is more efficient in hardware consumption. Our design uses only one-third LUTs and one-fourth DFFs compared with Ref. [23], and one-third LUTs and one-third DFFs compared with Ref. [43]. Although the RO-TDC in Ref. [42] has similar hardware consumption, its dead time is significant (maximum 602 ns). Although our GCO-TDC consumes slightly more logic resources than previously published GCO-TDCs [39]–[41], it is acceptable since the proposed TDC significantly improves both the resolution and linearity.

Although calibration methods like the bin-by-bin calibration [51], binwidth calibration [32] and mixed calibration methods [35] can improve linearity and precision, they all need manual calibration. To achieve automatic calibration, the gain and error calibration [45] and weighted calibration [46] methods are proposed. However, these designs only offer fixed resolutions. TDCs in Ref. [23], [43] and [44] provide flexible resolutions, but they all need manual configuration when resolution requirements change. To our knowledge, the VBCM is the first to achieve online resolution configuration and automatic calibration simultaneously in FPGA-TDCs. The C&C core serves 16 channels in this report but can serve more channels if required. Hence, the number of channels can be easily extended, and each channel's average logic resource consumption (with the C&C core) can be reduced.

We use Verilog to implement the proposed TDC. As the GCO-TDC's resolution and linearity are sensitive to placing and routing strategies, it requires a few constraints to guarantee that GCO's placements and routes are immobile. Firstly, we used Vivado Tcl commands "set_property BEL" and "set_property LOC" to place LUTs and DFFs [52] manually and then used commands "set_property LOCK_PINS" and "set_property FIXED_ROUTE" to lock LUTs' input pins and fix routing resources, respectively [52]. We can verify the "Input Shaper", GCO and sampling matrix by post-implementation simulations and the C&C core by behavior simulations. Although the design of the proposed TDC is slightly more complex than TDL-TDCs, it is acceptable since our TDC with automatic calibration is hardware cost-effective, highly linear and resolution-configurable.

## V. CONCLUSION

We proposed a new sampling structure, the sampling matrix, to enhance resolutions of GCO-TDCs. With this new structure, GCO-TDCs achieve excellent performances and low hardware consumption. We also proposed the VBCM to achieve automatic calibration and online resolution configuration. Besides, the hardware implementation of this method is detailed in this paper, and it is hardware-efficient through multiplexing critical components.

To evaluate our design, we implemented the proposed 16-channel TDC in UltraScale+, UltraScale and Virtex-7 FPGAs. Experimental results indicate that the proposed TDCs have competitive linearity and excellent uniformity. Due to online resolution configuration, they can have broader applications in TOF-LiDAR, PET-CT, or time-resolved spectroscopy (such as Raman spectroscopy). It can also be utilized as a TDC-core in prototype designs or commercial products, benefiting from automatic calibration and low resource consumption.

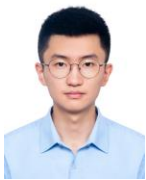
**Yu Wang** was born in Chongqing, China in 1995. He received the B.Eng. degree in measurement and control from the Harbin University of Science and Technology, in 2017, and the M.Eng. degree in electronics and communication engineering from Harbin Engineering University, in 2020. Since 2020, he has been working toward the Ph.D. degree founded by China Scholarship Council at the University of Strathclyde, Glasgow, U.K. His current research interests include FPGA-based mixed signal circuits.

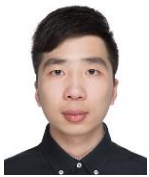
**Wujun Xie** was born in Hunan, China, in 1996. He received the B.Eng. degree in electronic and information engineering from the Hangzhou Dianzi University, Hangzhou, China, in 2017, and the M.S. degree in embedded systems from the University of Southampton, Southampton, U.K., in 2018. He has been working toward the Ph.D. degree at the University of Strathclyde, Glasgow, U.K since 2018.

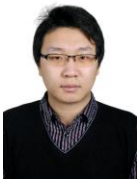
**Haochang Chen** was born in Xian China, in 1990. He received the M.S. degree in embedded digital systems from the University of Sussex, Brighton, U.K., in 2013, and the PhD degree from the University of Strathclyde, Glasgow, U.K., in 2020, funded by EPSRC, and then he joined the Fraunhofer UK Research Ltd as a researcher. His current research interests include FPGA-based high-precision time metrology systems for ranging and biomedical imaging applications.

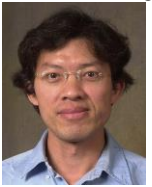
**David Day-Uei Li** received the Ph.D. degree in electrical engineering from the National Taiwan University, Taipei, Taiwan, in 2001. He then joined the Industrial Technology Research Institute, Taiwan, working on fast wireless and wireless communication chipsets. From 2007 to 2011, he worked at the University of Edinburgh on two European projects focusing on CMOS single-photon avalanche diode sensors and systems. He then took the lectureship in biomedical engineering at the University of Sussex, Brighton, in mid-2011, and in 2014 he joined the University of Strathclyde, Glasgow, as a Senior Lecturer. He has authored more than 90 journal and conference papers and holds 12 patents. His research exploits advanced sensor technologies to reveal low-light but fast biological phenomena.